\documentstyle[12pt]{article}
\newcommand{\rf}[1]{(\ref{#1})}
\newcommand{\bea}{\begin{eqnarray}}
\newcommand{\eea}{\end{eqnarray}}
\newcommand{\g}{\gamma}

\renewcommand{\b}{\beta}

\newcommand{\m}{\mu}

\newcommand{\sg}{\sigma}

\newcommand{\bm}{\bar{\m}}

\newcommand{\cT}{{\cal T}}

\newcommand{\cO}{{\cal O}}

\def\void{}
\def\labelmark{}

\newenvironment{formula}[1]{\def\labelname{#1}
\ifx\void\labelname\def\junk{\begin{displaymath}}
\else\def\junk{\begin{equation}\label{\labelname}}\fi\junk}%
{\ifx\void\labelname\def\junk{\end{displaymath}}
\else\def\junk{\end{equation}}\fi\junk\labelmark\def\labelname{}}

{\ifx\void\labelname\def\junk{\end{array}\end{displaymath}}
\else\def\junk{\end{array}\right.\end{equation}}
\fi\junk\labelmark\def\labelname{}\def\junk{}
}

\newcommand{\beq}{\begin{formula}}
\newcommand{\eeq}{\end{formula}}
\newcommand{\beqv}{\begin{formula}{}}

\begin{document}
\topmargin 0pt
\oddsidemargin 5mm
\headheight 0pt
\headsep 0pt
\topskip 9mm

\hfill    NBI-HE-94-02

\hfill January 1994

\begin{center}
\vspace{24pt}
{\large \bf A solvable 2d gravity model with
\mbox{$ \boldmath \gamma > 0$}~}\footnote{The authors acknowledge the support
from Nato Science Collaboration Grant no. 890138}

\vspace{24pt}

{\sl J. Ambj\o rn }

\vspace{6pt}

 The Niels Bohr Institute\\
Blegdamsvej 17, DK-2100 Copenhagen \O , Denmark\\

\vspace{12pt}

{\sl B. Durhuus}

\vspace{6pt}

Mathematics Institute\\
Universitetsparken 5, DK-2100 Copenhagen \O, Denmark\\

\vspace{12pt}

{\sl T. Jonsson}

\vspace{6pt}

Science Institute, University of Iceland, \\
Dunhaga 3, 107 Reykjavik, Iceland

\vspace{12pt}

\end{center}

\addtolength{\baselineskip}{0.20\baselineskip}
\vfill

\begin{center}
{\bf Abstract}
\end{center}

\vspace{12pt}

\noindent
We consider a model of discretized 2d gravity interacting with Ising
spins where phase boundaries are restricted to have minimal length
and show analytically that the critical exponent $\gamma= 1/3$
at the spin transition point.
The model captures the numerically observed behavior of standard multiple
Ising spins coupled to 2d gravity.

\vfill

\newpage

\section{Introduction}\label{sec-1}

Despite the progress in our understanding of 2d gravity coupled
to matter, both in the continuum description \cite{kpz,david2,dk}
and as a statistical model of random triangulations \cite{david,adf,kkm},
we have still no understanding  of the region which may be
most interesting, i.e. matter fields with central charge $c>1$
interacting with 2d gravity. One beautiful feature of 2d gravity coupled
to conformal matter theories with $c<1$ is that the influence of
matter on the geometry of the 2d universes only depends on the
central charge. This is expressed by the KPZ formula \cite{kpz,david2,dk}.
Recent numerical simulations suggest rather surprisingly that the
same is true at least for some range of $c>1$ \cite{at}.
It is often conjectured that $c>1$ is associated
with $\g >0$, where $\g$ denotes the string susceptibility.
In this letter we describe a solvable model with non-trivial critical
behavior for which $\g= 1/3$ and which explicitly realizes a
scenario for $c>1$ advocated in \cite{durhuus} (see also \cite{ambjorn}).
For other attempts to cross the $c=1$ barrier
we refer to \cite{burwick,msy}.

\section{The model}\label{sec-2}

A discretized model of 2d quantum gravity is defined by summing over
the triangulations one can obtain by gluing together equilateral
triangles along their links to form  closed surfaces,
and where one assigns the weight to each triangulation
dictated by Regge calculus. In the following we will restrict
ourselves to triangulations with spherical topology. This means
that we can ignore the Einstein-Hilbert term in the action. The
cosmological term is proportional to the
number of triangles so the partition function is given by:
\beq{*1}
Z(\m_) = \sum_{T \in \cT} \frac{1}{C_T}e^{-\m N_T} = \sum_N e^{-\m N}
\sum_{T \in \cT_N} \frac{1}{C_T}
\eeq
where $\cT$ denotes the class of triangulations of the sphere\footnote{
We find it convenient to consider only non-degenerate triangulations,
i.e. each triangle in the surface has three different vertices.
In terms of the dual $\phi^3$ graph it means that
we exclude tadpole graphs.}, $C_T$ a combinatoric factor
present for triangulations of closed surfaces, $N_T$ the number of triangles
in the triangulation $T$ and $\cT_N$ the subclass of $\cT$ whose
triangulations consist of $N$ triangles. Since
\beq{*2}
\sum_{T \in \cT_N} \frac{1}{C_T} =  e^{\m_0 N}\;N^{\g_0-3}\;(1+\cO (1/N))
\eeq
the model has a critical point $\m_0$ at which the continuum limit has
to be taken. It is known that $\g_0 = -1/2$.

It is easy to couple
matter fields to this discretized model. As an example one can
couple Ising spins to the geometry by placing one spin $\sg_i$ on each
triangle $i$. The partition function is now given by
\beq{*3}
Z(\m,\b) = \sum_N e^{-\m N} \sum_{T \in \cT_N} \frac{1}{C_T}
\sum_{\{\sg_i\}} \exp \left(\frac{\b}{2} \sum_{(ij)} (\sg_i \sg_j -1)\right)
\eeq
where $\sum_{(ij)}$ denotes the summation over all pairs of neighboring
triangles for a given triangulation $T$ and $\sum_{\{\sg_i\}}$ the
summation over the spin configurations on $T$.
We can couple multiple Ising spins to the discretized model in the
same way. For a fixed lattice such an extension would be trivial
since the models would not interact, but here they interact via
their backreaction on the geometry.

The partion function \rf{*3} depends on two coupling constants.
For each $\b$ there is a critical point $\m_c (\b)$ (and a $\g(\b)$)
where the continuum
limit has to be taken. In this way the possible candidates\footnote{The
mere existence of a critical point in the disretized model does of
course not ensure that there is an underlying continuum field theory.}
for a continuum theory will be labelled by the spin coupling $\b$.
In the case of a single Ising model one can explicitly solve the
model \cite{kazakov,bk} and the result is as follows: There is
a critical value of $\b$, $\b_c$,
above which there is spontaneous  magnetization
and below which the magnetization is zero. Away from $\b_c$ the
backreaction of the Ising spins on the geometry is effectively zero
($\g(\b) = \g_0$), while $\g (\b_c)=-1/3$,
in agreement with the KPZ formula for a $c=1/2$ conformal theory.
On a regular lattice the spin transition is a second order transition
and the corresponding conformal field theory has $c=1/2$.
On the dynamical lattice we see a corresponding transition at $\b_c$, but
the transition is modified by the coupling to gravity, and in addition
the backreaction of matter has modified the gravity theory.

For $c>1$, i.e. for more than two Ising models coupled to gravity,
no analytical solution of the model is known. However, numerical
simulations indicate the following:
\begin{itemize}
\item[1)] There is still a critical
point $\b_c$ below which there is no magnetization and above
which the system is magnetized. \cite{adjt,bj,ckr}
\item[2)] Above $\b_c$ the geometry seems to be that of pure
gravity. Below $\b_c$ there is a region where the situation is unclear,
especially for a large number of Ising spins,
but there seems to be a range of values of $\b$ for which the fractal
structure of the surface is very pronounced.
For sufficiently small $\b$ the geometry again is that of pure
2d gravity \cite{adjt,bj}.
\item[3)] For a sufficiently large number of Ising spins $\g(\b_c)$
will be positive \cite{bh,ajt,at}.
\item[4)] The geometry at $\b_c$ seems to be a function
of central charge $c$ of the matter system alone, at least for some
range of $c >1$. By this we mean that the
effective distribution of geometries is the same no matter
whether we consider $2N$ Ising spins at their critical point
or $N$ gaussian fields or any other conformal field theory with central
charge $c=N$, coupled to gravity \cite{at}.
\end{itemize}

In view of point 2) above it seems reasonable to attempt a description
of the model (for a large number of Ising copies) in a region around
$\b_c$ and for large $\b$ in terms of an effective model where baby
universes, i.e. parts of the surface connected to the rest by a
small loop, are completely magnetized and where a transition should
reflect the alignment of spins in different baby universes. Below we
shall define such a model (see also \cite{durhuus})
which can be solved explicitly by very simple means and show that it, indeed,
captures the features listed above except for the pure gravity
region for $\b$ small.

The model is obtained by restricting the summation in \rf{*3} over
Ising spin configurations as follows: Given a configuration
$\{\sg_i\}$ on a triangulation $T$ the corresponding spin clusters
consist of the maximal connected subsurfaces of $T$ on whose triangles
the spins are aligned and we shall require that the boundary components
of all spin clusters are minimal, i.e. of length 2. In other words, we
require all phase boundaries separating spins of opposite sign to be
loops of length 2. We shall henceforth indicate summation over
spin configurations with this property by $\sum'_{\{\sg_i\}}$.
Spin clusters with a boundary of length 2 are clearly the
excitations of the spins with lowest energy. Our summation
$\sum'_{\{\sg_i\}}$ represent an self-consistent
iteration of such spin excitations.

Thus the one-point (or one-loop) function  $G(\m,\b)$ is defined as
\beq{*4}
G (\m,\b) = \sum_{T \in \cT_1} e^{-\m N_T}{\sum_{\{\sg_i\}}}'
\exp \left(\frac{\b}{2} \sum_{(ij)} (\sg_i \sg_j -1)\right)
\eeq
where $\cT_1$ denotes the class of triangulations whose boundary is
a loop consisting of two (marked) links and for later convenience
the spins on the two boundary triangles are fixed to one.
Similarly, one defines
$n$-point functions that are essentially derivatives w.r.t. $\m$ of
the one-point function.

We note that $G(\m,\b)$ is well defined and finite in a region of
the $(\m,\b)$-plane that contains the domain of definition for
the full Ising model coupled to 2d gravity and is contained in the
half plane $\m > \m_0$, where $\m_0$ is the critical point of
pure 2d gravity, whose one-point function is given by
\beq{*5}
G_0 (\m) = \sum_{T \in \cT_1} e^{-\m N_T},
\eeq
More specifically, it follows by standard arguments that there
exists a critical curve $(\m_c'(\b),\b)$ with
\beq{*6}
\m_0 \leq \m'_c (\b) \leq \m_c(\b)
\eeq
such that $G(\m,\b)$ is analytic in $\m,\b$ on the right of this curve.
It is also easy to show that $\m'_c(\b) \to \m_0$ as $\b \to \infty$.

The susceptibility $\chi(\m,\b)$ is defined as
\beq{*7}
\chi(\m,\b) = - \frac{\partial G}{\partial \m}
\eeq
and the string susceptibility exponent $\g (\b)$ is given by
\beq{*8}
\chi(\m,\b) = f(\m-\m'_c(\b),\b)+ \frac{c(\b)}{(\m -\m'_c(\b))^{\g(\b)}} +
{\rm less~~singular~~terms},
\eeq
where $f$ is an analytic function.

The corresponding quantities in  pure 2d gravity will be denoted by
$\chi_0(\m)$ and $\g_0$. In particular $\g_0$ is given by eq. \rf{*2}
and, as mentioned above, equals $-1/2$.

\section{Critical behavior}\label{sec-3}

In order to determine the critical behavior of the model we
establish a self-consistency equation as follows. Let $T \in \cT_1$
be a triangulation with a spin configuration contributing to
$G(\m,\b)$. Each phase boundary (of length 2) separates a baby
universe from the rest of the triangulation and clearly any two
such baby universes are either disjoint or one contains the other.
Thus we may in a unique way cut off maximal baby universes bounded by
phase boundaries and close up the corrresponding boundary loops of
length 2 in the remaining part of $T$ to obtain a triangulation
$\bar{T} \in \cT_1$ (with the same boundary as $T$) all of whose
spins are aligned. Conversely, we obtain a $T \in \cT_1$ and a
spin configuration $\{\sg_i\}$ contributing to $G(\m,\b)$ by
starting with a $\bar{T} \in \cT_1$ with all
spins aligned, cutting it open along a set of links and finally gluing
on baby universes with appropriate spin configurations along the
corresponding loops of length 2 such that the spin on the two
boundary triangles are oppositely oriented to those of $\bar{T}$.

Using this decomposition procedure and summing first over baby
universes with spin configurations we obtain
\beq{*9}
G(\m,\b) = \sum_{\bar{T} \in \cT_1} e^{-\m N_{\bar{T}}}
\left( 1+ e^{-2\b} G(\m,\b)\right)^{L_{\bar{T}}-2}
\eeq
where $L_{\bar{T}}$ denotes the number of links in $\bar{T}$,
i.e. $L_{\bar{T}} = 1+\frac{3}{2}N_{\bar{T}}$. In eq. \rf{*9}
the factor $e^{-2\b}$ represents the coupling of a baby
universe across the phase boundary to the rest of the surface and
the factor 1 in the parentheses originates from the empty baby universe.
We can now reexpress eq. \rf{*9}
as\footnote{Strictly speaking we have added a factor
$(1+e^{-2\b} G(\m,\b))$ on the rhs of \rf{*9} since
$L_T-2= \frac{3}{2}N_T-1$. However, all arguments presented in the
follows are valid even if we did not include this factor. We have chosen to
add it in order to simplify the formulas.}
\beq{*10}
G(\m,\b) = \sum_{T\in \cT_1} e^{-\bm N_T} = G_0(\bm),~~~~~~~~
\bm=\m-\frac{3}{2} \log \left(1+ e^{-2\b}G(\m,\b)\right).
\eeq
Note that the last equation can be written as
\beq{*11}
\m = \bm + \frac{3}{2} \log \left( 1 + e^{-2\b} G_0(\bm)\right)
\eeq
which expresses $\m$ in terms of known functions of $\bm$ and $\b$
since pure gravity can be solved.

{}From eq. \rf{*10} and \rf{*11}  we get
\bea
\chi(\m,\b) &=& \chi_0 (\bm) \frac{\partial \bm}{\partial \m}\label{*12}\\
\frac{\partial \bm}{\partial \m} &=&
\frac{e^{2\b} + G_0 (\bm)}{e^{2\b}-
(\frac{3}{2} \chi_0 (\bm) -G_0(\bm))}. \label{*13}
\eea

Since the string susceptibility exponent $\g_0 = -1/2 < 0$
in the case of pure
gravity both $G_0(\m_0)$ and $\chi_0(\m_0)$ are finite.
This implies that there exists a $\b_c$, given by
\beq{*13a}
e^{2\b_c}= \frac{3}{2}\chi_0 (\m_0) -G_0(\m_0)
\eeq
such that the denominator in \rf{*13} is different from
zero for all $\m \geq  \m'_c(\b)$ provided $\b >\b_c$ .
By differentiating $\chi(\m,\b)$ $n$  times with respect to
$\m$ we know from \rf{*8} that  $\chi^{(n)} (\m,\b)$ will be
singular for $\m \to \m'_c (\b)$ provided $n$ is sufficiently large.
Performing the same differentiation on the rhs of \rf{*12}, using
again \rf{*13}, we see that the only chance for a singular
behavior is that $\bm (\m'_c(\b) = \m_0$ and in this case the
leading singularity has to come
from $\chi_0^{(n)}(\bm)$. We conclude:
\beq{*14}
\bm(\m'_c(\b)) = \m_0 ~~~{\rm and}~~~~\g(\b) = \g_0~~~~{\rm for}~~~~
\b > \b_0.
\eeq
This is the phase where the model is magnetized, where
the spin fluctuations are small and where the geometry of the
surfaces is not  affected by the spins.

The number $\b_c$ (given by \rf{*13a})
is characterized by being the largest $\b$  for which
$\partial \m/\partial \bm$ equals zero for some $\bm \geq \m_0$,
i.e. (by \rf{*13}) the largest $\b$ for which the equation
\beq{*15}
e^{2\b} = \frac{3}{2} \chi_0(\bm) - G_0 (\bm) ~~
(=\sum_{T \in \cT_1} (3N_T/2-1)e^{-\bm N_T})
\eeq
has a solution for $\bm \geq \m_0$. If we define $\bm_c (\b)$ by
\beq{*16}
\left. \frac{\partial \m}{\partial \bm} \right|_{\bm_c(\b)} = 0
\eeq
for $\b \leq \b_c$,  then $\bm_c(\b)$ obviously  solves \rf{*15} and we have
\beq{*17}
\bm_c(\b_c) = \m_0,~~~~~~\bm_c(\b) > \m_0 ~~~{\rm for}~~~\b < \b_c.
\eeq

Let us now assume that $\b < \b_c$.
If we use eq. \rf{*12} and \rf{*13} this implies that $\chi(\m,\b)$
will be singular for  $\bm\to \bm_c(\b)$ due to the vanishing denominator
on the lhs of eq. \rf{*13}.
In fact, since $\bm_c(\b) > \m_0$ both $\chi_0 (\bm)$
and $G_0(\bm)$ will be regular around $\bm_c(\b)$ and we can Taylor expand
the lhs of \rf{*12}:
\beq{*18}
\chi(\m,\b) \sim \frac{c}{\bm -\bm_c(\b)} \sim
\frac{ \tilde{c}}{\sqrt{\m -\m'_c(\b)}}
\eeq
To derive the last equation we have used \rf{*13} and
\rf{*16} which tell us that
\beq{*19}
\m= \m'_c(\b) + const (\bm-\bm_c(\b))^2 +\cdots.
\eeq
We conclude that $\g(\b) = 1/2$ for $\b < \b_c$. In this phase
baby universes are  dominant. Effectively we have branched
polymers  and the total magnetization of the system is zero \cite{adjt}.

Let us finally consider the system at the critical point $\b_c$.
This point is characterized by the fact that $\bm_c(\b_c)$ coincides
with $\m_0$. Although the singularity of $\chi(\m,\b_c)$ for
$\m \to \m'_c(\b_c)$ is still dominated by the zero of
$\partial \m/\partial \bm$ we can no longer Taylor expand $\m (\bm,\b_c)$
around $\bm_c(\b_c)$ ($=\m_0$) since the functions in \rf{*13} are singular
in $\m_0$. On the other hand we can use the known singular behavior of
the pure gravity functions $G_0$ and $\chi_0$
at $\m_0$ to deduce from \rf{*13}, remembering that $\g_0 < 0$,
\beq{*20}
\frac{\partial \m}{\partial \bm} \sim (\bm-\bm_c(\b_c))^{-\g_0},
{}~~~~~{\rm i.e.}~~~\m-\m'_c(\b_c) \sim (\bm-\bm_c(\b_c))^{-\g_0+1}.
\eeq
{}From \rf{*12} we finally get, using \rf{*20}
\beq{*21}
\chi(\m,\b_c) \sim \frac{c}{(\bm-\bm_c(\b_c))^{-\g_0}} \sim
\frac{\tilde{c}}{(\m-\m'_c(\b_c))^{-\g_0/(-\g_0+1)}},
\eeq
and we have derived the remarkable relation:
\beq{*22}
\g(\b_c) = \frac{-\g_0}{-\g_0+1} =\frac{1}{3}.
\eeq

\section{Conclusion}\label{sec-5}

We have seen above that our toy model has two generic phases,
a magnetized phase for large $\b$ where the distribution of
geometries coincides with that of pure gravity, as one naively
would expect for large $\b$, and a phase where $\g(\b) =1/2$.
The exponent $\g=1/2$ is that of branched polymers and the
total magnetization of such a system is zero due to the
linear structure of branched polymers \cite{adjt}. At the critical point
$\b_c$ for magnetization we have $\g(\b_c)=1/3$.

Our model shows how the interaction between matter and geometry
can lead to a string susceptibility $\g >0$. It agrees remarkably
well qualitatively with numerical simulation of
multiple Ising models except for a region of small  $\b$ which
shrinks with increasing multiplicity of the Ising models.
The model is closely related to the matrix models studied in \cite{matrix}
and  to the $c \to \infty$ limit  of multiple
Ising models studied in \cite{wexler}.
However, our approach has the virtue of being simple and
avoids any use of matrix models by working directly with the
spin excitations on the surfaces. Moreover, it highlights the general nature
of eqs. \rf{*12} and \rf{*13}.

It is a natural assumption that the full multiple spin model will have
logarithmic corrections to $\g=1/3$ (if that is indeed the
correct exponent) and that these will increase as
the central charge $c$ decreases from infinity. There is even some
numerical support for this conjecture \cite{at}.

\vspace{24pt}

\noindent  T.J. acknowledge the support from the Fulbright Foundation and would
like to express his thanks to for the hospitality at ITP and the
Department of Mathematics at UCSB.

\vspace{24pt}

\addtolength{\baselineskip}{-0.20\baselineskip}


\begin{thebibliography}{99}
\bibitem{kpz}V. Knizhnik, A. Polyakov and A. Zamolodchikov, Mod.Phys.Lett.
A3 (1988) 819.
\bibitem{david2}F. David, Mod.Phys.Lett. A3 (1988) 1651.
\bibitem{dk}J. Distler and H. Kawai,  Nucl.Phys. B321 (1989) 509.
\bibitem{david}F. David, Nucl. Phys. B 257 (1985) 45;
F. David, Nucl. Phys. B257 (1985) 543.
\bibitem{adf}J. Ambj\o rn, B. Durhuus and J. Fr\"{o}hlich, Nucl. Phys. B 257
 (1985) 433; Nucl.Phys. B270 (1986) 457; Nucl. Phys.  B275 (1986) 161.
\bibitem{kkm}V. A. Kazakov, I. K. Kostov and A. A. Migdal, Phys. Lett. 157B
 (1985) 295;  D. Boulatov, V. A. Kazakov, I. K. Kostov and A. A. Migdal,
Nucl. Phys. B275 (1986) 641.
\bibitem{at}J. Ambj\o rn and G. Thorleifsson, {\it A universal
fractal structure of 2d quantum gravity for $c >1$}, NBI-HE-93-64,
hep-th/9312157, to appear in Phys.Lett.B.
\bibitem{durhuus}B. Durhuus, {\it Multi-spin systems on a random
triangulated surface}, preprint.
\bibitem{ambjorn}J. Ambj\o rn, {\it Barriers in quantum gravity},
NBI-HE-93-31, to appear in {\it Strings 93}, Wold Scientific Publishers.
\bibitem{burwick}T.T. Burwick,
{\it All order quantum gravity in two dimensions},
SLAC-PUB-6274, hep-th/9307074.
\bibitem{msy}M. Martellini, M. Spreafico and K. Yoshida, {\it A continuum
approach to 2d quantum gravity for $c >1$},  ROME-961-1993.
\bibitem{kazakov}V.A. Kazakov, Phys.Lett. A119 (1986) 140.
\bibitem{bk}D. Boulatov and V.A. Kazakov, Phys.Lett. B186 (1987) 379.
\bibitem{bh}E. Brezin an S. Hikami, Phys.Lett. B283 (1992) 203;
S. Hikami, Phys.Lett. B305 (1993) 327.
\bibitem{adjt}J. Ambjorn, B. Durhuus, T. Jonsson and G. Thorleifsson,
Nucl.Phys. {\bf B398} (1993) 568; \\
G. Thorleifsson, Nucl.Phys. B30 (Proc.Suppl.) (1993) 787.
\bibitem{bj} C.F. Baillie and D.A. Johnston, Phys.Lett. B286 (1992) 44;
Mod.Phys.Lett.  A7 (1992) 1519.
\bibitem{ckr}S. M. Catterall, J.B. Kogut and R.L. Renken, Nucl.Phys. B292
(1992) 277; Phys.Rev. D45 (1992) 2957.
\bibitem{ajt}J. Ambj\o rn, S. Jain and G. Thorleifsson, Phys.Lett. B307
(1993) 34.
\bibitem{matrix}S.R. Das, A. Dhar, A.M. Sengupta and S.R. Wadia,
Mod.Phys.Lett. A5 (1990) 1041. \\
G.P. Korchemsky, Phys.Lett. {\bf B296} (1992) 323;
{\it Matrix models perturbed by higher curvature terms}, UPRF-92-334.\\
L. Alvarez-Gaume, J.L.F. Barbon, and C. Crnkovic,
Nucl.Phys. {\bf B394} (1993) 383.
\bibitem{wexler}M. Wexler, Phys.Lett B315 (1993) 67; Mod.Phys.Lett. A8
(1993) 2703.

\end{thebibliography}
\end{document}